\documentclass{iaus260}
\usepackage{graphicx}
\pubyear{2009}
\volume{260}  %% insert here IAU Symposium No.
\pagerange{10--17}
\jname{The R\^ole of Astronomy in Society and Culture}
\editors{D. Valls-Gabaud \& A. Boksenberg, eds.}
\title[Radio quiet, please!]                   %% Give here short title %%
{Radio quiet, please! -- protecting radio astronomy from interference}  %% Give here full title  %%

\author[W. van Driel]           %% Give here short author list %%
{W. van Driel$^{1,2}$}        %% Give here full author list %%

\affiliation{$^1$GEPI, Observatoire de Paris, CNRS, Universit\'e Paris Diderot, \\
             5 place Jules Janssen, 92190 Meudon, France  \\
             email: {\tt wim.vandriel@obspm.fr} \\[\affilskip]
             $^2$Scientific Committee on Frequency Allocations for Radio Astronomy \\
             and Space Science (IUCAF) \\
                 }

\begin{document}
\maketitle

\begin{abstract}
The radio spectrum is a finite and increasingly precious resource for astronomical research, as well as for other spectrum users. Keeping the frequency bands used for radio astronomy as free as possible of unwanted Radio Frequency Interference (RFI) is crucial. The aim of spectrum management, one of the tools used towards achieving this goal, includes setting regulatory limits on RFI levels emitted by other spectrum users into the radio astronomy frequency bands. This involves discussions with regulatory bodies and other spectrum users at several levels -- national, regional and worldwide.
The global framework for spectrum management is set by the Radio Regulations of the International Telecommunication Union, which has defined that interference is detrimental to radio astronomy if it increases the uncertainty of a measurement by 10\%. The Radio Regulations are revised every three to four years, a process in which four organisations representing the interests of the radio astronomical community in matters of spectrum management (IUCAF, CORF, CRAF and RAFCAP) participate actively. The current interests and activities of these four organisations range from preserving what has been achieved through regulatory measures, to looking far into the future of high frequency use and giant radio telescope use.
    %% Add here a maximum of 10 keywords
\end{abstract}

\firstsection % If your document starts with a section,
\section{Introduction}
When people see the nowadays standard ``Did you switch off your mobile phone?'' sign flash by on the screen before the start of a movie in a theatre, they will understand and cooperate -- everybody knows the nuisance value of a loud phone conversation one metre away. When people see the same sign in a hospital they usually cooperate as well, but they may do so out of the mistaken understanding that they are in a place where silence is required, rather than the fact that their phones can severely disturb the medical scanners further down the corridor. Finding such a sign in the middle of nowhere, with a notice attached that there is a radio telescope operating 50 km away, would presumably merely baffle them. 

We astronomers have an educational role in making others -- active radio spectrum users as well as the general public -- aware of the high sensitivity of radio telescopes and the consequent need for protection of the Radio Astronomy Service. The International Year of Astronomy provides us with a unique opportunity in this respect.

The radio window (wavelength range$\sim$300 $\mu$m-30 m, or frequencies $\sim$10 MHz-1000 GHz -- Fig. 1) is the only other domain in the electromagnetic spectrum, besides the optical/infrared window, which can be observed with ground-based telescopes. It provides us with a unique look on the Universe that allows measurements and discoveries that can be made nowhere else in the spectrum.

\begin{figure}[h]
% \vspace*{-2.0 cm}
\begin{center}
 \includegraphics[width=11cm]{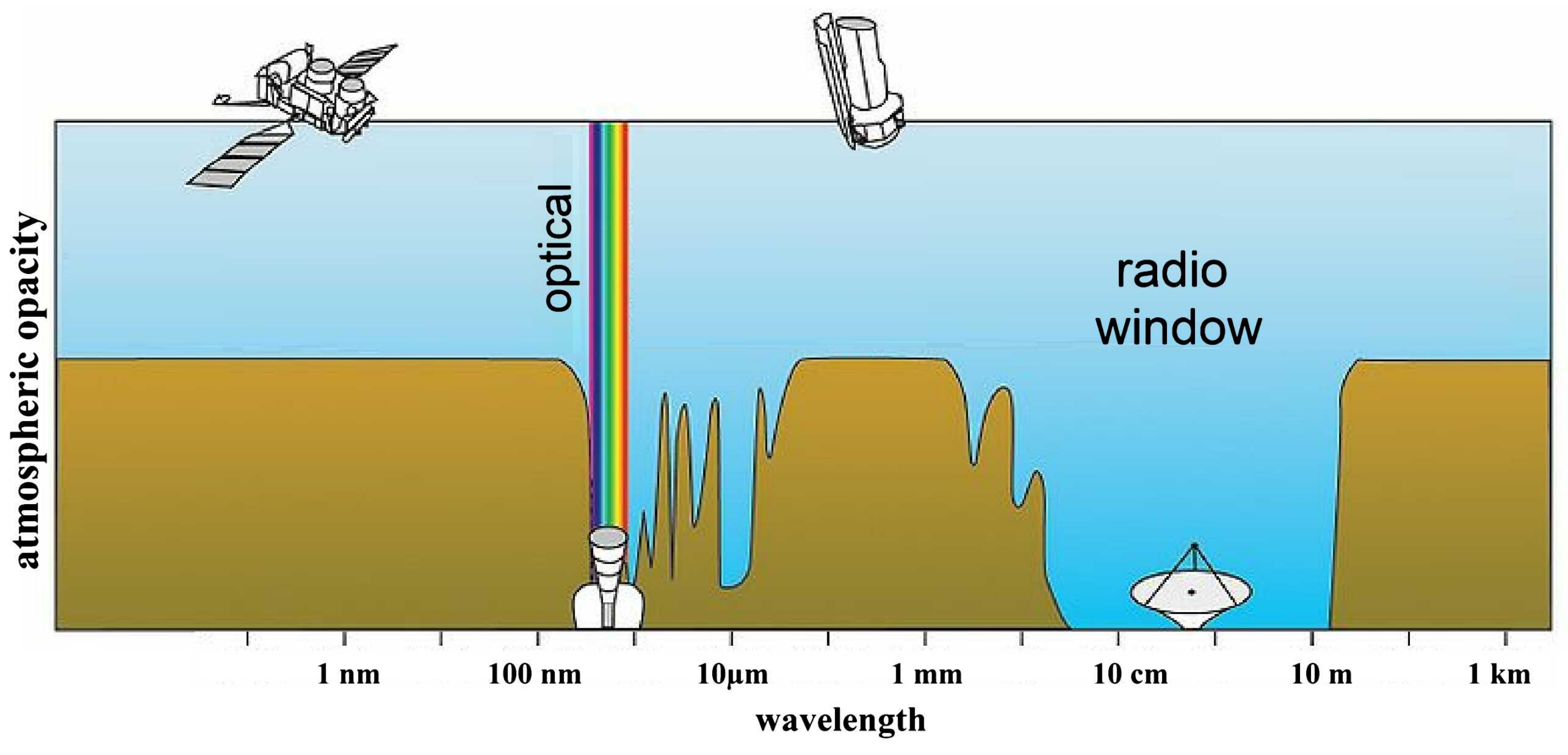} 
% \vspace*{-1.0 cm}
 \caption{The electromagnetic spectrum, with the two spectral windows in which astronomical observations are possible with ground-based instruments: the optical/infrared window and the radio window. The radio window spans a factor 100,000 in wavelength, from 0.3 mm to 30 m.}
   \label{fig1}
\end{center}
\end{figure}

\section{Radio Frequency Interference: unwanted emissions}

Looking for RFI, he found Astronomy - Karl Jansky (1932): 
Unwanted Radio Frequency Interference (RFI) has been with us since the beginning -- in fact, it lies at the very origin of the detection of radio waves of celestial origin.
Investigating static for long-distance radio telephone service he found ``a faint steady hiss of unknown origin'' at a frequency of 20 MHz (=$\lambda$14.5 m), whose moment of peak intensity shifted by about 4 minutes per day, which a local school teacher told him meant it had to be of celestial origin: the Galactic Centre.

Looking for Astronomy, he found RFI - Grote Reber (1940's):
While mapping the radio emission from the Milky Way with a home-built parabolic radio telescope his data were full of spikes due to car ignitions. He found that the RFI in his data was manageable, however: the intermediate time scales of the radio signal from the Milky Way was different from that of the car ignitions and the slowly drifting baselines due to receiver gain instability (Fig. 2).

Radio telescopes keep getting larger, but not necessarily more sensitive to RFI:
Radio telescope sensitivity has doubled every 3 years -- so far, by an impressive factor of 
100,000 in the 60 years since Grote Reber -- and the active (i.e., emitting) users of the radio spectrum keep getting busier. Although this may sound like a scenario that has to lead to an inevitable catastrophe for radio astronomy, in practice this has not turned out to be the case throughout the radio spectrum, as most RFI enters through telescope sidelobes (which have not become more sensitive with time), from directions (far) away from where the telescope is pointed, and from (far) outside the frequency band in which the astronomical observations are made (Fig. 4).

\begin{figure}[t]
% \vspace*{-2.0 cm}
\begin{center}
 \includegraphics[width=11cm]{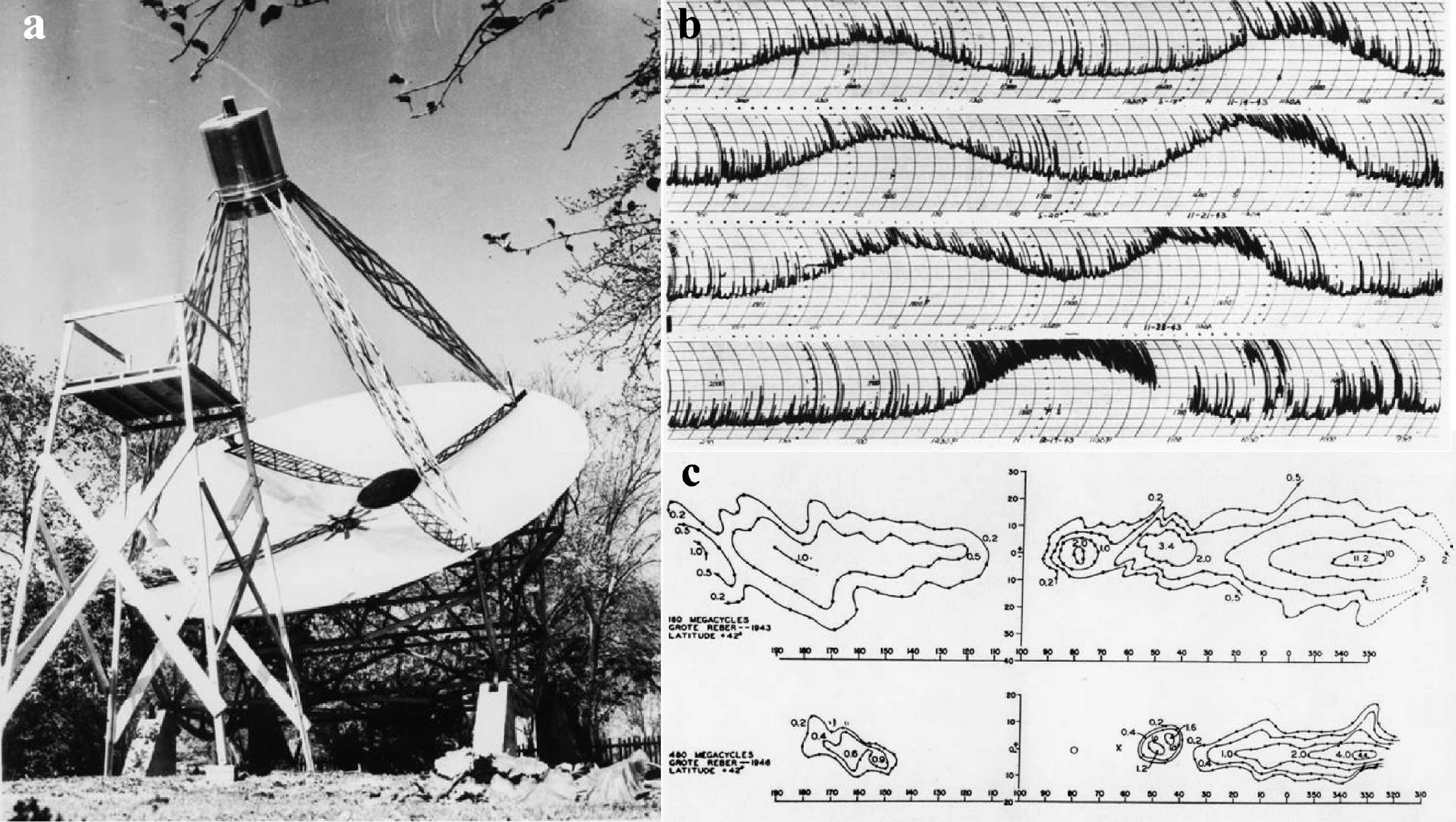} 
% \vspace*{-1.0 cm}
 \caption{When in the 1940's the pioneer Grote Reber mapped the radio sky at frequencies of 160 and 480 MHz with his wooden parabolic radio telescope (a) he found lots of spiky Radio Frequency Interference (mainly car ignitions) on top of celestial radio signals (b), which he could disentangle due to their different time behaviour, to produce his radio sky maps (c).}
   \label{fig2}
\end{center}
\end{figure}

RFI -- how bad can it get (an informal scale -- see Table 1): Its negative impact on the quality and 
reliability of radio astronomical observations can range from mildly inconvenient to smoke coming from 
radio telescope receivers (and astronomers), cf. Fig. 3 -- following IUCAF (2004). 
The official regulatory definition of ``detrimental interference'' is given below.

\begin{table}
  \begin{center}
  \caption{Levels of RFI -- an informal scale}
  \label{tab1}
 {\scriptsize
  \begin{tabular}{ll}\hline 
{\bf RFI level} & {\bf its effect}  \\ \hline
Manageable   &   much weaker than desired signal  \\
Inconvenient &   need to repeat observations  \\
Embarrassing &   weak, but leads to fake ``detection''  \\
Distressing  &   multiple failures  \\
Obliterative &   pointless to observe further  \\
Destructive  &   permanent damage to receiver \\ \hline
 \end{tabular}
  }
 \end{center}
\end{table}

\begin{figure}[h]
% \vspace*{-2.0 cm}
\begin{center}
 \includegraphics[width=10.5cm]{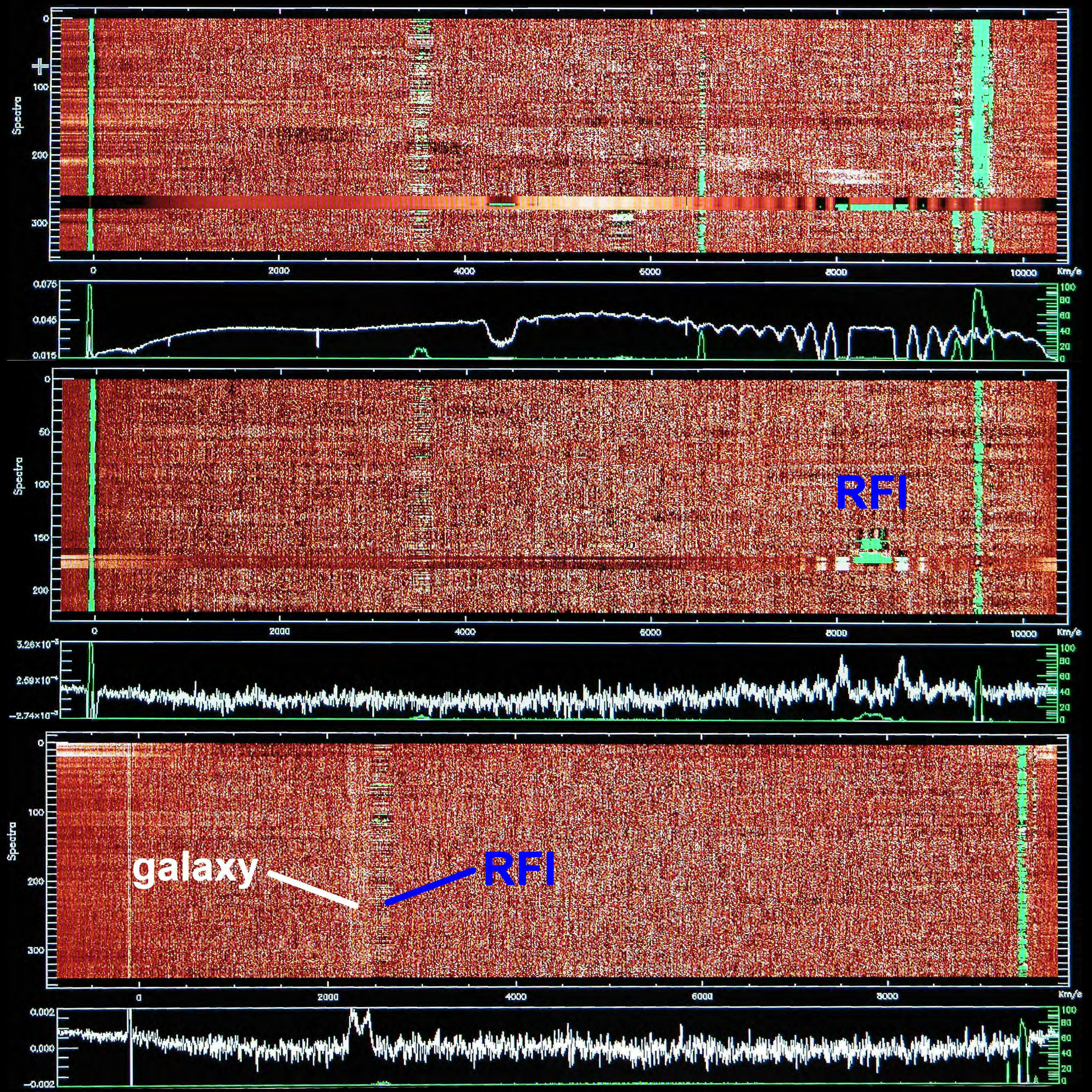} 
% \vspace*{-1.0 cm}
 \caption{Three levels of RFI in 21cm H{\footnotesize I} line spectra taken with a single-dish radio 
telescope for a galaxy redshift survey. Each panel shows a so-called waterfall display of the data, 
with time as vertical axis and frequency (expressed as Doppler-shifted radial velocity in km/s) 
as horizontal axis; the green data were flagged as strongly time-variably RFI and removed from the time-averaged spectrum shown as the white plot underneath each waterfall display: Top panel: inconvenient (RFI too strong during part of the observation, need to repeat observation), middle: embarrassing (remaining satellite RFI around 8200 km/s mimics a double-horned galaxy H{\footnotesize  I} profile), and bottom: manageable (the RFI on the flank of the H{\footnotesize I} profile of a galaxy at 2300 km/s is faint enough).}
   \label{fig3}
\end{center}
\end{figure}

How can we keep/get RFI out of our data?:
Given the many ways in which RFI can get into our data this is a complex situation without 
a ``magic bullet'' solution, i.e., no single technique exists that can address all possible scenarios. 
Methods to avoid and/or minimize the impact of RFI mitigation can be divided into two broad categories: 
regulatory (spectrum management) and technical (RFI mitigation).
Regulatory methods involve putting regulatory (legal) limits on RFI into radio astronomy frequency bands, and technical methods are aimed at removing RFI signals from radio astronomy observations.
The scope of the present review is limited to spectrum management aspects.

\begin{figure}[h]
% \vspace*{-2.0 cm}
\begin{center}
 \includegraphics[width=10cm]{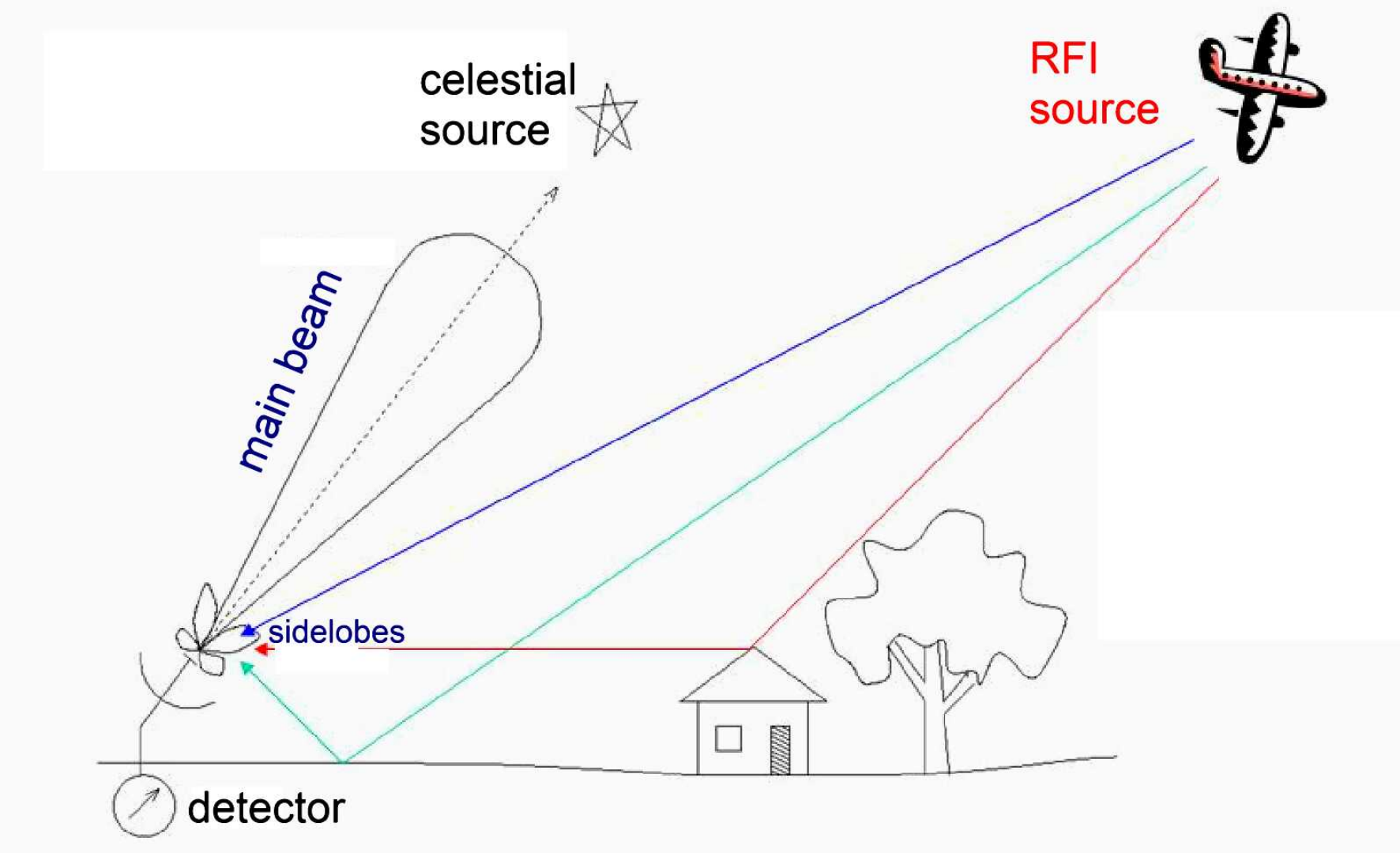} 
% \vspace*{-1.0 cm}
 \caption{A sketch of the various ways in which RFI from a radio transmitter can enter into radio astronomical observations -- through various paths, both direct and reflected (adapted from Cohen, in IUCAF 2004).}
   \label{fig4}
\end{center}
\end{figure}

\section{Spectrum management: regulatory solutions}
The radio spectrum is a finite and increasingly precious resource for astronomical research, as well as for other spectrum users -- a lot of who want to use the spectrum to make a lot of money, and for whom there is little economic incentive to implement methods (filters, etc.) that reduce RFI into radio astronomy bands.

Spectrum management can be defined as the task of accommodating all competing radio services and systems within the finite range of the radio frequency spectrum, e.g., by allocating frequency bands, and defining limits on RFI. 
Its underlying principle is ``Prevention is Better than a Cure'': putting regulatory limits on RFI that can be emitted into the radio astronomy frequency bands will make it easier to deal with it.

Regulating the electromagnetic smog: over a hundred new satellites are launched each year, and new applications of radio and electronics keep appearing, but few other radio services notice the rising levels of radio pollution, as they do not perform measurements down to the sensitivity levels of large radio telescopes, and they can usually crank up their transmitting power to stay above the RFI -- an option radio astronomy does not have, as we cannot increase the transmission power of our celestial radio sources... 

What price radio astronomy?: while national spectrum management Administrations tend to prefer deregulation and spectrum pricing, where billions in the local currency are involved in auctioning off relative small portions of the radio spectrum, the literal question ``What price the radio astronomy bands?'' arises. For an exploratory report that gives examples of attempts to quantify the significant economic and societal value of scientific use of the radio spectrum see RSPG (2006), in which it is stated that 
`Scientific usage of spectrum has considerable societal weight and economic value. It might be difficult to quantify the benefits of scientific use as they can relate to society as a whole, may be difficult to foresee and maybe be realised over very long periods of time.'

\subsection{Our regulatory workhorse to define and limit RFI: Rec 769} 
Called for short ``rec seven-six-nine'', its formal name is ITU-R Recommendation RA.769 on ``Protection criteria used for radio astronomical measurements''. It gives threshold levels for RFI detrimental to radio astronomy observations, i.e., interference that introduces a change in amplitude (voltage) at the output of a receiver equal to 1/10 of the rms noise, in a total-power measurement -- in other words, RFI that increases the uncertainty of a radio astronomy measurements by 10\%, i.e., which reduces the effective integration time by 20\% -- see the ITU-R Radio Astronomy Handbook for more details.

The Recommendation gives detrimental RFI threshold levels for both continuum and spectral line measurements with single-dish telescopes (Total Power), linked interferometers (like Westerbork),  and VLBI. Active service operators often consider our detrimental RFI levels ``unrealistically low'', and open to negotiation, but they are not used to detecting faint celestial radio sources. For us radio astronomers, Rec 769 is literally ``The Limit'', and its levels are non-negotiable -- accepting RFI that is 10 dB above the Rec 769 threshold level means we will loose the use of that band in practice. 

\subsection{Spectrum Management problem space} 
In Figure 5, which illustrates the complex ``problem space'' of Spectrum Management, there are three axes (passive services, active services, Administrations) and spheres with three different radii (national, regional, global). This indicates that the practical regulatory issues confronting us have a very wide scope and require interactions at quite different levels -- they can vary from dealing with local authorities on a planned mobile phone tower that is too close to a radio observatory to working at a global level towards defining RFI limits on unwanted emissions from satellites. \\

\vspace{-2mm}
{\it Active and passive Services, and Administrations: }\\
\vspace{-3mm}

In regulatory terms, the Radio Astronomy Service is considered a ``passive'' (i.e., non-emitting) service, meaning that we do not emit electromagnetic radiation ourselves, but build large radio telescopes in order to detect faint natural radio sources. The active services do emit radiation, and their goals are often orthogonal to our own: the RFI they generate can potentially degrade the quality of radio astronomy observations to levels which are unacceptable to us, but the practical means to avoid this (e.g., through installing extra filters in satellites, which adds to their launch weight and cost) may be considered an ``undue burden'' by them. The role of the Administrations is to ensure the equitable use of the radio spectrum, which often includes playing the role of an impartial referee between the active and passive radio services. \\ 

\vspace{-2mm}
{\it National, regional and global spheres: } \\
\vspace{-3mm}

Each sovereign state has, in some way or other, its own spectrum management Administration with the mandate to use all means possible to facilitate and regulate (enforce) radiocommunication in that country. Their mandates and terms of reference are usually defined by national telecommunication laws, which also include a national frequency allocation table, which is the national articulation of the ITU Radio Regulations. 

Between the detailed frequency planning necessary for national Administrations and the broad global framework established by the ITU, there has always been a need for regional coordination, which is being provided by APT, CEPT and CITEL. 

\begin{figure}[t]
%\vspace*{-0.3 cm}
\begin{center}
 \includegraphics[width=10.5cm]{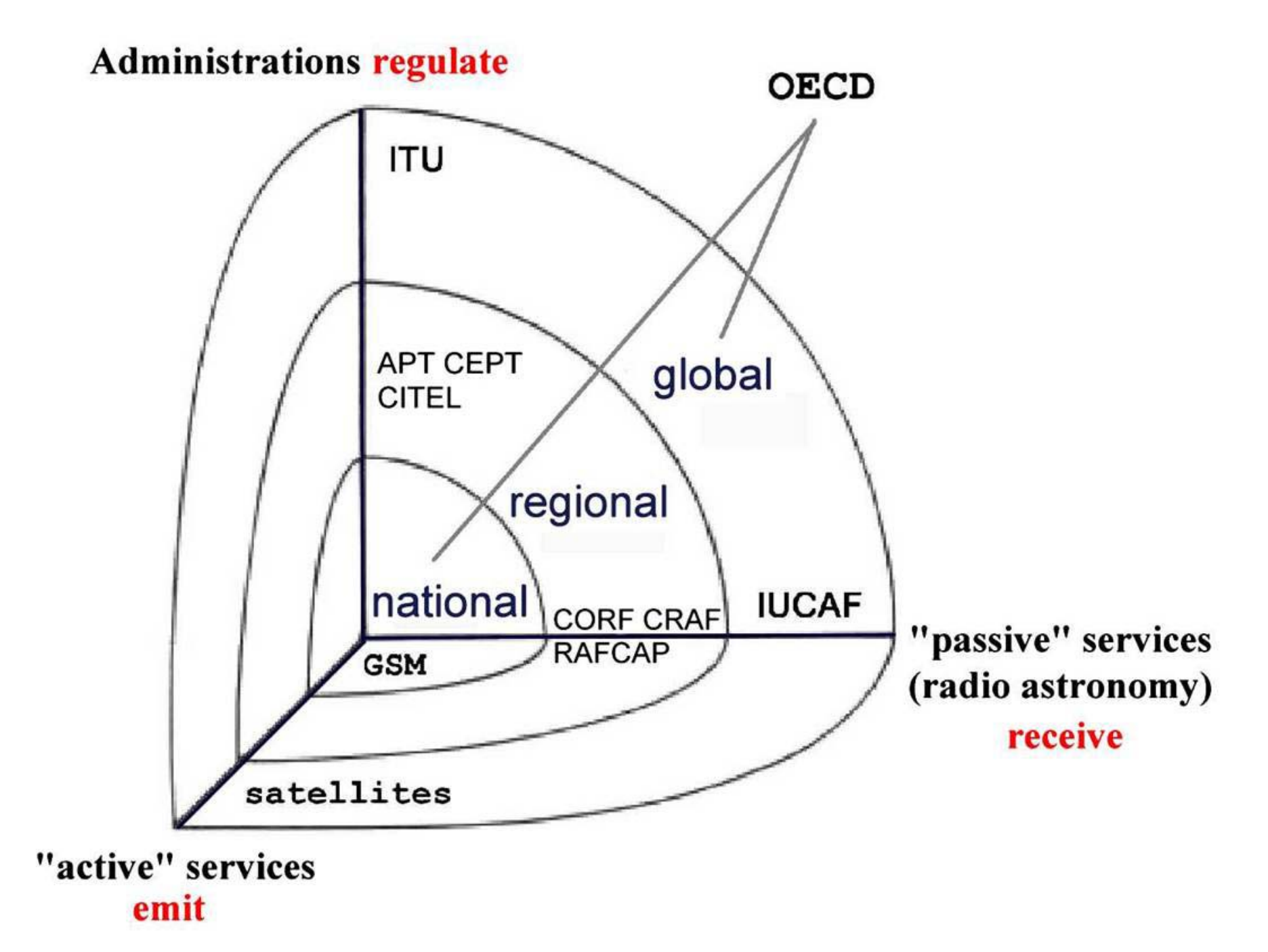} 
% \vspace*{-1.0 cm}
 \caption{The complex ``problem space'' of radio spectrum management, with three orthogonal axes (passive services, active services, Administrations) and three spheres of different radii (national, regional, global).}
   \label{fig5}
\end{center}
\end{figure}

The international Administrative cooperation body that co-ordinates spectrum management at the global level is the International Telecommunication Union, ITU (www.itu.int). 
The global framework for radio spectrum management is provided by the Radio Regulations of the ITU, which have international treaty status and thus are binding for all members of the ITU. They provide rules to national Administrations that allow them to regulate equitable access to the radio spectrum for all entities requiring frequency allocations: telecommunication industry, safety services, aeronautical services, various scientific and hobby uses, etc. The Radio Regulations contain the international Frequency Allocation Table, and, e.g., rules for the use, operation and coordination of frequencies - including limits on RFI into radio astronomy bands. The Radio Regulations are updated every three to four years, during an ITU World Radio Communication Conference (WRC) -- a four week-long event involving 3000 participants, including 15 astronomers.

\section{Radio Quiet Zones}
Optimizing the protection of a radio observatory from RFI requires a package of measures to deal with the many difference aspects of interference: global regulatory protection, strong national and local protection, and efficient RFI mitigation techniques.

Radio observatories are located in remote areas, in order to avoid RFI from active spectrum users and radio noise produced in industrial or residential areas. Most observatories are surrounded by a Radio Quiet Zone (RQZ), which was set up using state or national laws. A Radio Quiet Zone typically consists of two zones (Cohen et al. 2005): an exclusion zone in which all radio emissions are prohibited, with restrictions on housing and industrial developments, and a larger (up to 100+ km radius) coordination zone where the power of radio transmissions is limited to levels that are based on compatibility studies with radio astronomical observations.

Compatibility with radio astronomy in Radio Quiet Zones is usually based on detrimental RFI levels defined in ITU-R Recommendation RA.769 for frequency bands allocated to radio astronomy, whose underlying principles can also apply anywhere outside these frequency bands in the recently defined RQZs for potential sites for the planned giant Square Kilometre Array radio telescope.

The electronic and electrical equipment that is used at radio observatories themselves can also potentially interfere with our own observations, and it is therefore a common sight to see computers and correlators at observatories enveloped in a Faraday cage, or to have rooms or even entire buildings shielded against radio noise that may leak out. 

\section{Radio astronomers and spectrum management}
In spectrum management matters, the interests of the radio astronomy community is represented by a worldwide organisation, IUCAF, and three regional organisations, CORF (USA), CRAF (Europe), and RAFCAP (Asia-Pacific). Most are Sector Members of the ITU, which enables them to participate directly in studies and deliberations in various ITU fora. The four organisation work together in close coordination.

IUCAF (www.iucaf.org), the Scientific Committee on Frequency Allocations for Radio Astronomy and Space Science, is sponsored by three Scientific Unions, COSPAR, the IAU, and URSI. Its brief is to study and coordinate the requirements of radio frequency allocations for passive (i.e., non-emitting) radio sciences, such as radio astronomy, space research and remote sensing, in order to make these requirements known to the national and international bodies that allocate frequencies. IUCAF operates as a standing inter-disciplinary committee under the auspices of ICSU, the International Council for Science. It organises or sponsors series of Summer Schools on spectrum managements for radio astronomy and Workshops on RFI mitigation techniques.

CORF (www7.nationalacademies.org/corf/), the Committee on Radio Frequencies of the National Academy of Science, represents US radio astronomy interests. It is actively involved in the coordination and consensus seeking between the entities representing government and non-government radio spectrum interests in the United States. The Electromagnetic Spectrum Manager of the National Science Foundation (NSF) is charged with securing access to the spectrum for the government science enterprise, mostly radio telescopes operated by the US national centres (NRAO and NAIC). CORF has participated in the Handbook of Frequency Allocations and Spectrum Protection for Scientific Uses (CORF et al. 2007).

CRAF (www.craf.eu), the Expert Committee on Radio Astronomy Frequencies of the European Science Foundation, represents the European radio astronomy community. 
Although, especially from the outside, ``Europe'' is commonly regarded as equivalent to those countries assembled in the European Union, for spectrum management matters Europe covers a considerably larger territory: the 48 countries of the CEPT, the European Conference of Postal and Telecommunications Administrations (www.cept.org). 
Its members represent the radio astronomical observatories of 19 CEPT countries, the European VLBI Network (EVN), the Joint Institute for VLBI in Europe (JIVE) and 3 other multi-national organisations (EISCAT, ESA, and IRAM). CRAF employs a full time pan-European radio astronomy Spectrum Manager. It has published the CRAF Handbook for Radio Astronomy (CRAF 2005) and the CRAF Handbook for Frequency Management (CRAF 2002). Furthermore, CRAF regularly publishes a Newsletter, which is available on its website. 

RAFCAP (www.atnf.csiro.au/rafcap/), the Radio Astronomy Frequency Committee in the Asia-Pacific region, represents our interests in the region which primarily comprises countries in from South and East Asia, Oceania and the Pacific islands, while excluding the Americas. Organisations in the Asia- Pacific region face special challenges in coping with the very diverse cultures and languages of the different nations. It is modelled on CRAF. The main forum for RAFCAP activities is the APT (Asia-Pacific Telecommunity), and more specifically its preparations for the World Radiocommunication Conferences of the ITU.

The interests and activities of these four organisations range from preserving what has been achieved through regulatory measures or mitigation techniques, to looking far into the future of high frequency use and giant radio telescope use. Current priorities, which will certainly keep them busy through the next years, include the use of satellite down-links close in frequency to the radio astronomy bands, the coordination of the operation in shared bands of radio observatories and powerful transmissions from downward-looking satellite radars, the possible detrimental effects of ultra-wide band (UWB) transmissions and high-frequency power line communications (HF-PLC) on all passive services, the scientific use of the 275 to 3000 GHz frequency range, and studies on the operational conditions that will allow the successful operation of future giant radio telescopes.

\end{document}